# Understanding magnetotransport signatures in networks of connected permalloy nanowires

B. L. Le[1], J.-S. Park[1], J. Sklenar[1], G.-W. Chern[2], C. Nisoli[3], J. Watts[4], M. Manno[4], D. W. Rench[5], N. Samarth[5], C. Leighton[4], P. Schiffer[1]

[1]Department of Physics and Frederick Seitz Materials Research Laboratory, University of Illinois at Urbana-Champaign, Urbana, Illinois 61801, USA

[2]Department of Physics, University of Virginia, Charlottesville, Virginia 22904, USA

[3]Theoretical Division, Los Alamos National Laboratory, Los Alamos, New Mexico 87545, USA

[4]Department of Chemical Engineering and Materials Science, University of Minnesota, Minneapolis, Minnesota 55455, USA

[5]Department of Physics and Materials Research Institute, Pennsylvania State University, University Park, Pennsylvania 16802, USA

## Abstract

The change in electrical resistance associated with the application of an external magnetic field is known as the magnetoresistance (MR). The measured MR is quite complex in the class of connected networks of single-domain ferromagnetic nanowires, known as 'artificial spin ice', due to the geometrically-induced collective behavior of the nanowire moments. We have conducted a thorough experimental study of the MR of a connected honeycomb artificial spin ice, and we present a simulation methodology for understanding the detailed behavior of this complex correlated magnetic system. Our results demonstrate that the behavior, even at low magnetic fields, can be well-described only by including significant contributions from the vertices at which the legs meet, opening the door to new geometrically-induced MR phenomena.

Magnetoresistance (MR) plays a central role in studies of magnetism [1, 2], and MR effects are particularly interesting in nanostructures, where size constraints can alter the fundamental electrical transport behavior [3]. A prominent example of such nanostructures, connected artificial spin ice, consists of mesoscopic networks of single-domain ferromagnetic nanowires arranged in lattice geometries designed to emulate the frustrated behavior of spin ice [4, 5, 6]. In zero magnetic field, the moments of the nanowire legs are effective Ising spins aligned with their long axes, directed toward and away from the vertices of the lattice. Local interactions result in so-called “ice rules” that govern the number of moments pointing into and out of each vertex to minimize the local magnetostatic energy. Artificial spin ice structures have proven to be exemplary systems in which to study frustration [5,6] and have been noted for their technological potential as both a reconfigurable metamaterial and as a memory storage medium [7, 8, 9, 10, 11, 12]. Connected artificial spin ice in particular has been studied extensively in the last decade [13, 14, 15, 16, 17, 18, 19, 20, 21, 22,23,24, 25]. Tanaka *et al*. and other workers, observed sharp features in the magnetoresistance of connected artificial spin ice, and associated them with reversal events that maintain the ice rules of the system [13, 14, 15]. An explicit understanding of the data, however, has not yet been achieved beyond qualitative attribution to changes in the magnetization profile in combination with anisotropic magnetoresistance (AMR), a property of ferromagnetic metals associated with changes in resistivity as a function of the angle between the magnetization and the local current density [26, 27].

We have studied MR in artificial spin ice, combining experimental data with a micromagnetic-based transport model that gives us microscopic understanding of the observed sharp features in both the longitudinal and transverse magnetoresistance. We find that the MR behavior of artificial spin ice systems can be explained through a complex interplay of AMR and the collective magnetization response of the frustrated network, and that the physics of the system involves significant deviations from a simple spin ice model of Ising moments even at low magnetic fields. Specifically, we find that the vertices where nanowires intersect can be critical in determining the observed MR behavior of the entire system. While previous studies of magnetotransport in different geometries of ferromagnetic structures demonstrated that domain wall configurations would be important to transport measurements [28, 29, 30, 31, 32, 33, 34], here we are able to understand and quantitatively replicate details of magnetotransport considering both the collective magnetic structure of this frustrated system and the resulting complex electric field configurations throughout the system. The success of our methodology indicates that a wide range of new magnetotransport effects associated with nanoscale geometry can be understood in both artificial spin ice and other ferromagnetic nanoscale systems, opening possibilities for new devices and novel magnetoresistive physical phenomena.

We studied permalloy ($Ni_{81}Fe_{19}$) networks of nanowires, which we label as "legs" connected at vertices (Fig. 1d), patterned into a Hall-bar geometry with current leads spanning the width of the sample (Fig. 1a). Magnetic force microscopy imaging at zero field (Fig. 1b) confirmed the single domain nature of the nanowire legs of the networks. Applying an ac current, we measured both $V_{||}$, the longitudinal voltage, and $V_{\perp}$, the transverse voltage (Fig. 1c), and thus determine the longitudinal and transverse resistances $R_{||}$ and $R_{\perp}$. The in-plane magnetic field was at an angle θ to the nominal current direction denoted by θ. Experimental details are given in the Supplementary Sections 1 and 4 [35].

All data shown are from samples in the armchair geometry seen in Fig. 1a, where one third of the wires are parallel to the current. Measurements on a 90°-rotated lattice in the zigzag geometry, where one third of the wires are perpendicular to the current, are qualitatively consistent with the results discussed below (Supplemental Section 1 [35]). For all data shown, the sample dimensions were consistent, and the individual nanowire legs of the network were approximately 800 nm by 75 nm in-plane and 25 nm thick, with the vertex regions having approximate lateral dimension of 100 nm.

We plot our magnetoresistance data for field sweeps that start from +10000 Oe and sweep down to -10000 Oe, and then back to +10000 Oe. The maximum fields are sufficient to saturate the magnetization. Selected longitudinal resistance data are displayed in Fig. 2a, with corresponding transverse data shown in Fig. 2c. Field sweeps for a second armchair geometry device, and for other θ, can be found in Supplemental Section 2 [35]. The transverse data has an offset associated with slight longitudinal misalignment of the leads (less than ~50 nm). For all θ, we observed the broad parabolic background and field reversal symmetry expected for AMR [16]. Clear sharp features in the field sweep data are associated with changes in the magnetization as the nanowire leg moments collectively flip from being aligned with the magnetic field at the start of the sweep, to obeying the ice rules near zero field, and then becoming aligned again with the field at the end of the sweep.

In Fig. 3, we plot the transverse magnetoresistance response at a given field strength as a function of angle (in 5° increments). This composite angular plot was assembled from downward field sweep data. The high field transverse data have the expected symmetry for the transverse component of AMR, known as the planar Hall effect, with extrema at 45° and 135°. This confirms that the Ising character of the [26, 27] nanowire leg moments is suppressed in a strong external field, *i.e*., the applied field rotates the moments away from the wire axes. It is then natural to ask whether an Ising-ice model can describe the transport at low field. Since shape anisotropy should completely align nanowire leg moments with the wire axes at zero field, the legs at zero field should not result in any transverse magnetoresistance because of the functional form of the planar Hall effect [26, 27]. Surprisingly, the zero field remanent resistance state of the structure is also highly sensitive to the angle at which the field is swept, but with a completely different functional form consisting of three different plateaus with steps near 30°, 90° and 150°. This implies that the vertices, *i.e.*, the regions at which the nanowire legs of the structure meet, are playing a significant role in the measured transverse resistance. While the vertex magnetization profiles are naturally determined by the neighboring leg magnetization, the data indicate that a simple ice model considering only the legs of the network is insufficient to explain the magnetotransport of this system even at low fields.

The changes in resistance as a function of field was also minimized at $\theta = 0°$ and $\theta = 90°$, and we see intriguing behavior at those angles. In Fig. 4a, we show field sweeps for angles very close to $\theta = 90°$, where small changes in angle result in drastic differences in MR. For $\theta > 90°$, starting from negative field saturation, the resistance steadily increased, dropped sharply at 400 Oe, and then steadily increased until a jump at 1700 Oe. For $\theta < 90°$, we observed inverted behavior: the resistance steadily decreased, jumped sharply at 400 Oe, and then steadily decreased until a drop at 1700 Oe. This effect is repeated in zigzag-orientation networks for θ = 0° (Supplemental Section 1 [35]), for which the orientation of the magnetic field with respect to the lattice was the same.

To better understand the origins of MR behavior in this system, we developed an approach combining micromagnetic modeling with the phenomenology of AMR. Our transverse magnetoresistance data in Fig. 3 demonstrated the necessity of such modeling. We first obtained the magnetization profile $\mathbf{m}(x, y)$ of the structure using the GPU-based MuMax3 package [36], starting with the geometry of an SEM image of an experimental device to approximate the edge roughness. We used $\mathbf{m}(x, y)$ to calculate the longitudinal and transverse MR via computation of the electric field associated with AMR, as given by

$$\mathbf{E} = \rho_0 \, \mathbf{J} + \Delta\rho(\hat{\mathbf{m}} \cdot \mathbf{J})\hat{\mathbf{m}} \qquad (\mathbf{1})$$

where $\mathbf{J}$ is the electric current vector, $\hat{\mathbf{m}}$ is a unit vector in the direction of the magnetic moment, $\rho_0$ is the isotropic bulk resistivity, and $\Delta\rho$ is the anisotropic magneto-resistivity [26, 27].

For a given magnetization profile $\mathbf{m}(x, y)$, we adopted a perturbation approach appropriate for $\Delta\rho$ small relative to the total resistance, and a reduced sample of 17 hexagons, as shown in Fig. 5. We used a simplified current distribution such that $\mathbf{J}^{(0)} = J\,\hat{\mathbf{e}}_x$ (along the *x* direction) for the vertex regions and the horizontal legs, whereas $\mathbf{J}^{(0)} = (J/2)\,\hat{\mathbf{e}}_{1,2}$ for the remaining non-horizontal legs. Next, we use this 0$^{th}$-order current distribution and Eq. (1) to compute the first-order correction to the electric field: $\frac{\mathbf{E}^{(1)}}{\rho_0} = \frac{\Delta\rho}{\rho_0}\left(\hat{\mathbf{m}} \cdot \mathbf{J}^{(0)}\right)\hat{\mathbf{m}}$. A $\Delta\rho/\rho_0$ value of 0.05 was used as appropriate for permalloy [26, 27], acting here simply as a scaling factor. By taking the line integral of the first-order electric field, $\Delta V^{(1)} = \int_C \mathbf{E}^{(1)} \cdot d\boldsymbol{l}$ over the appropriate path *C*, we calculated both the longitudinal and transverse resistance by integrating the field and dividing by the current. Paths chosen mimicked the placement of the leads used in the experiment and are detailed in the Supplemental Section 3 [35]. This methodology is quite different from previous modeling of AMR in ferromagnetic nanostructures that relied on adding together the resistance from cells of the structures [33,34], and this new method allows for more complex structures and simulation of transverse resistances.

The results of this modeling are shown adjacent to our experimental data in Fig. 2 and 4. The agreement of the experimental longitudinal data (Fig. 2a) and the simulated longitudinal response (Fig. 2b) strongly corroborate this approach, capturing the parabolic background and qualitatively reproducing the observed features. The experimental transverse data (Fig. 2c) and the simulated transverse response (Fig. 2d) similarly show good agreement for most angles.

Fig. 4 compares the experimental and simulated transverse data for a small subset of angles around 90°. Notably, the modeling was able to capture both the overall qualitative features and the inversion of the plot features. As the field was increased from -4000 Oe to 500 Oe, the MR gradually rose ($\theta > 90°$) or fell ($\theta < 90°$) as the nanowire leg moments aligned with their axes near zero field, and the sharp feature at approximately 500 Oe was due to the magnetization reversal of the ±60° nanowire leg moments. The simulations also indicate that the drastic change with field angle is associated with the field being nearly perpendicular to the one third of the nanowires; a slight offset in angle forces the magnetization of those wires to all align in one direction at the highest field.

The close agreement between modeling and experimental data demonstrates that our technique successfully captures the physics of MR in this system, and the few angles that did not show good agreement (e.g., the transverse data for θ = 60°) we attribute to a failure of the first term perturbation approach, likely arising from the simplified current distribution, combined with the smaller size of the simulation lattice (compared to experimental samples). Another possibility is that at certain angles our simulation, which is at effectively zero-temperature, is not fully representative of our room temperature experiment [25] , suggesting a possible avenue for further study.

We hypothesized from our experimental data that the vertices are critical to the transverse MR, especially at field angles near ±90°, and our simulations allow a direct test of this conclusion. To demonstrate the impact of the vertex regions on the transverse MR, we separate the 90.2° simulated MR trace in Fig. 4b into two parts: the contribution to the MR from the nanowire legs and the contribution to the MR from the vertices (defined as triangular regions between adjacent legs as in Fig. 1d). This plot makes it clear that, over the experimental field range, the transverse MR originated mainlyfrom the vertex portions. We can qualitatively explain this fact based on a symmetry argument; the field near $90^{o}$ is symmetric with respect to the ±60° nanowires. When the electric field is integrated along the +60° leg, there is cancellation with a corresponding -60° leg. As can be seen in the Supplemental Section 3 [35], the vertex contribution is smaller for other angles of applied field, but it is again appreciable for the field applied at 0°, where the degree of symmetry for the structure is also high.

To provide a microscopic picture of the vertex contributions to the MR, we show a snapshot of our simulations in Fig. 5. Here, the system was initially magnetized with a - 4000 Oe field applied at 90.2º. The snapshot we show is at 800 Oe, the reversal point of the ±60° nanowires, where some, but not all, of those nanowire leg moments have reversed. Fig. 5a is a map of the magnetization, and Fig. 5b is the corresponding map of the y-component of the electric field. The polarity of the electric field for the nanowire legs is independent of whether or not they have reversed magnetization.

Fig. 5c is an expanded region of the magnetization map showing six vertices. The vertices have two possible magnetization profiles, depending on whether the adjoining nanowire leg moments have reversed. The three vertices on the right are adjacent to nanowire legs whose moments have reversed, with the leg magnetization pointing upward. The three vertices on the left are adjacent to wires whose moments have not reversed, with magnetization pointing downward. The corresponding map of the y-component of the electric field is shown in Fig. 5d, showing how the different vertex magnetization states result in a difference in the transverse voltage. We can similarly understand the three zero-field transverse resistance plateaus in Fig. 3, with transitions near 30°, 90°, and 150°. In these directions the field is perpendicular to approximately one-third of the nanowires. For field sweeps on either side of these angles, the zero-field remanent magnetization configuration changes, resulting in different vertex magnetization profiles and transverse resistances.

We emphasize that the transverse resistance has a many-body origin, because the nonzero transverse voltage in the small field regime arises from vertices obeying the ice rules. This is demonstrated by the 1-in-2-out and 2-in-1-out vertices in Fig. 5, where the minority spins are on the ±60° nanowires. For such configurations, there is a net magnetization vector pointing along the ±60° directions in the vertex region, which in turn gives rise to an electric field pointing in the same direction according to Eq. (1). This off-axis electric field associated with the ice-rule obeying vertices is the microscopic source of the electric field that yields the non-zero transverse resistance. This realization opens the possibility of designing reconfigurable magneto-resistance devices based on artificial spin ice.

Our results demonstrate how the complex magnetotransport of artificial spin ice networks can be understood through appropriate modelling. The unexpected contributions of vertex regions, and the concomitant extreme sensitivity of the behavior to field angle at certain field orientations, both suggest the possibility of new phenomena associated with the magnetoresistance of creative geometries, even in simple ferromagnetic metals. Furthermore, our methodology of integrating the electric field associated with AMR allows precise modelling of these effects in a wide range of nanostructures. Given the unusual physics inspired by the geometric freedom of artificial spin ice [37,38] and other creative structures enabled by modern lithography, exploitations of similar effects are likely to enable novel physical phenomena and device applications.

**Acknowledgments**

The authors acknowledge Jarrett Moyer and Paul Lammert for useful discussions. This project was funded by the US Department of Energy, Office of Basic Energy Sciences, Materials Sciences and Engineering Division under grant no. DE-SC0010778.

Work at the University of Minnesota was supported by the NSF MRSEC under award DMR-1420013, as well as by DMR-1507048. CN's work is carried out under the auspices of the NNSA of the U.S. DoE at LANL under Contract No. DE-AC52-06NA25396 and financed by DoE at the LANL IMS.

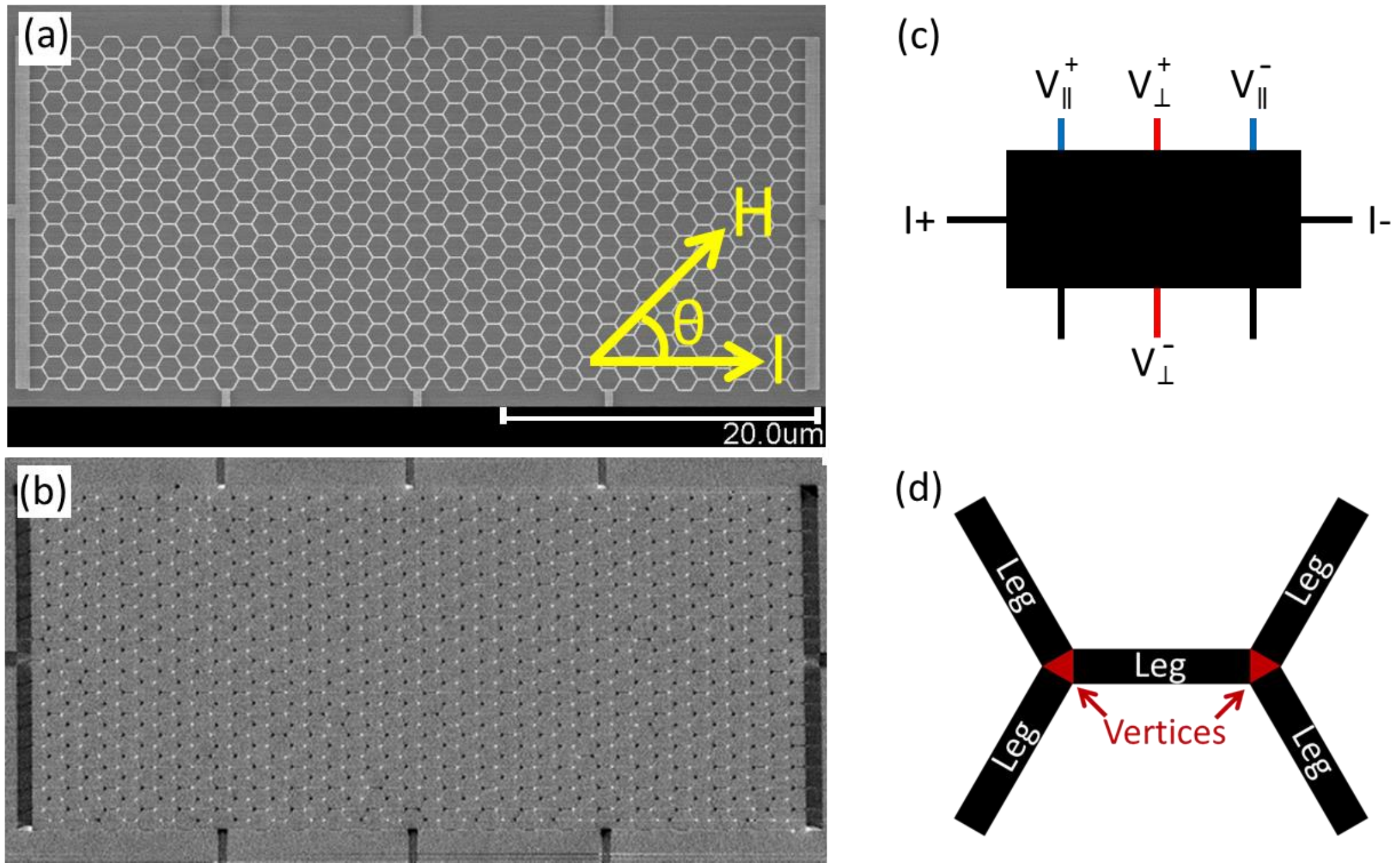

**Figure 1:** (a) SEM image of an armchair orientation connected kagome artificial spin ice lattice. An external magnetic field, H, could be applied along any in-plane direction, denoted by the angle θ between the field direction and the nominal current direction, I. (b) Corresponding MFM image. The black and white dots at the vertices are domain walls, indicative of the Ising-like behavior of the individual nanowires. (c) A schematic illustration of the lead arrangement used for transport studies. Large connective pads on each end supply an excitation current, while the thin nanowire leads along the long axis were used for voltage measurement. (d) A schematic illustration of the nanowire leg and vertex regions.

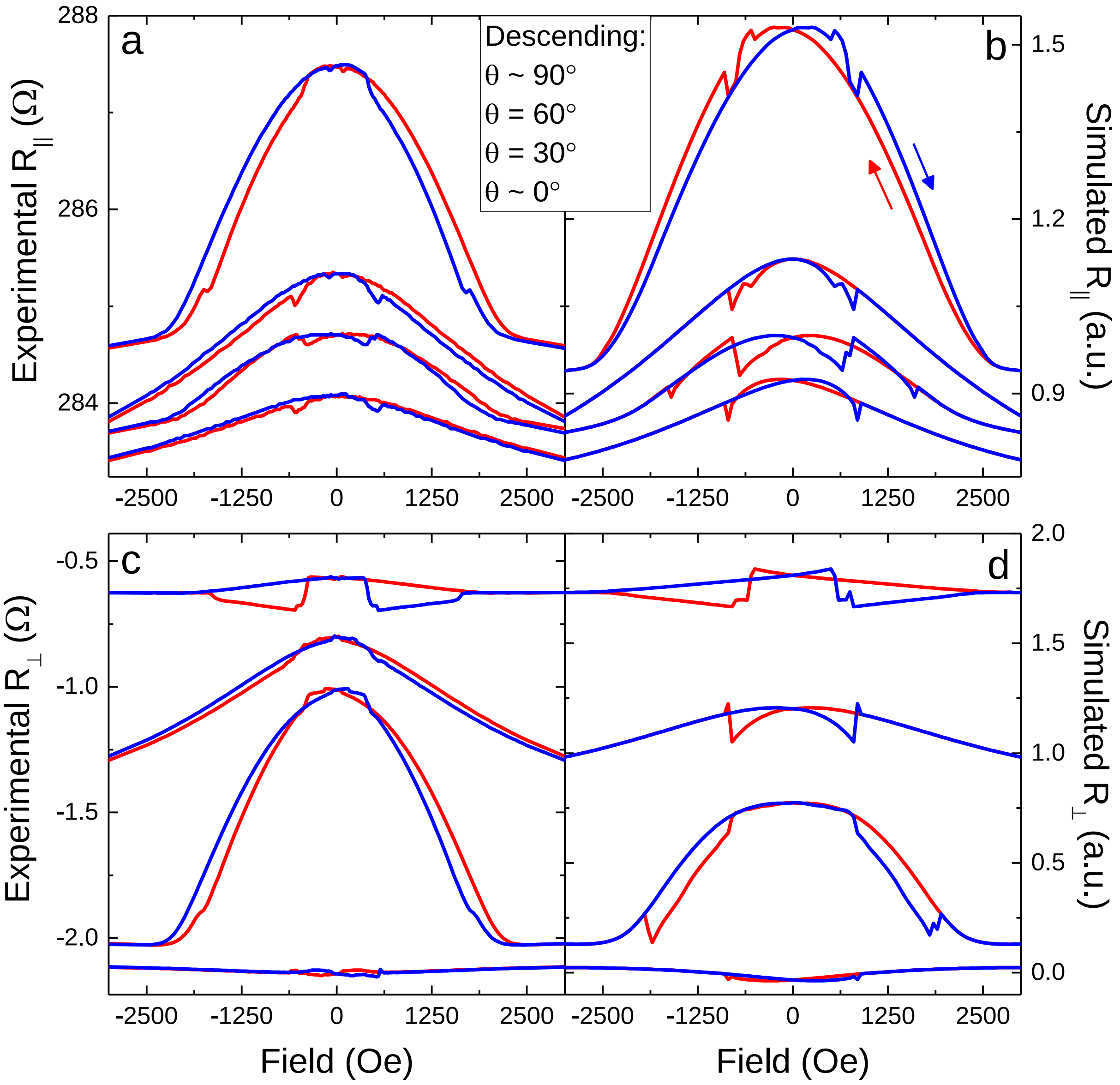

**Figure 2**: Experimental longitudinal (a) and transverse (c) magnetoresistance data, and corresponding simulated longitudinal (b) and transverse (d) magnetoresistance data. The down field sweeps (red) and up field sweeps (blue) are symmetric under field inversion. For viewing ease, all data except for θ = 0° have been vertically offset. Simulated data are for applied field angles at 90.2°, 60°, 30°, and 0.2°.

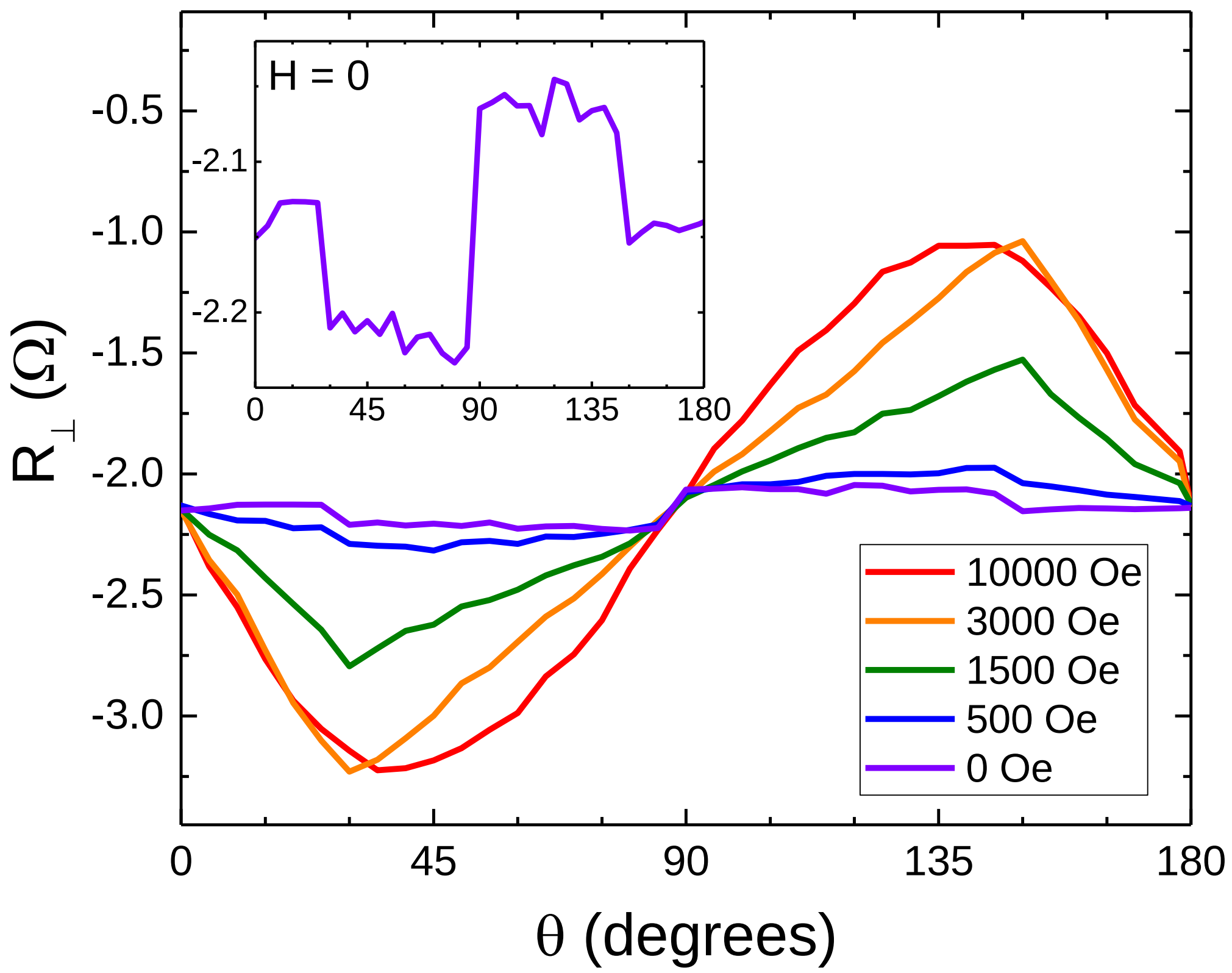

**Figure 3**: Pseudo-angular tracking of transverse resistance values for identical field strengths at varying θ. For each curve, the connected data were taken from sweeps of the magnetic field magnitude at different angles. The high-field data reveal the expected symmetry of the planar Hall effect, with extrema at 45 degrees and 135 degrees. Inset: Zero field transverse resistance values as a function of θ. The remanent zero-field resistance value depends on the angle at which the field was applied, reflecting the effect of the vertex regions.

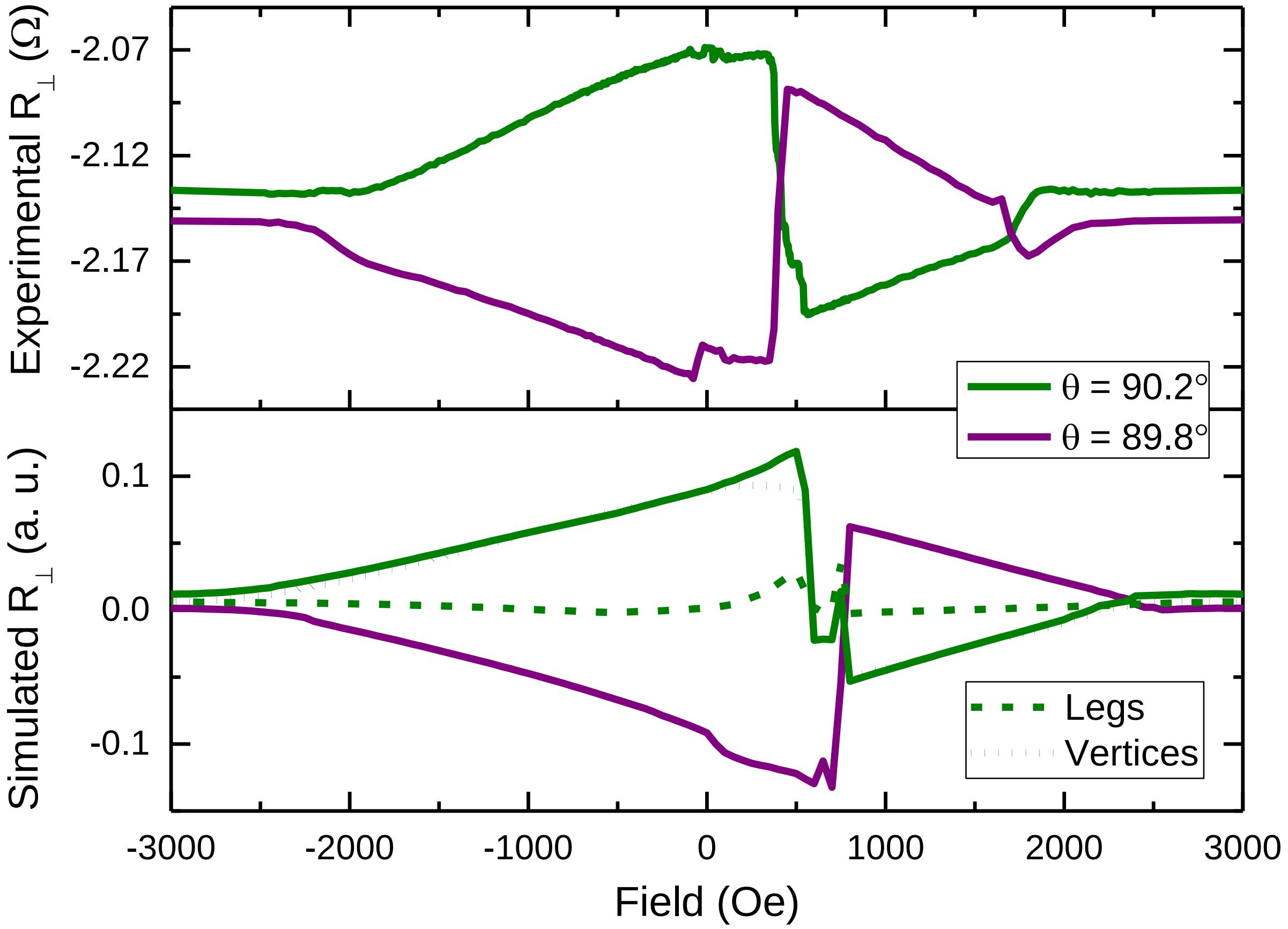

**Figure 4**: Transverse magnetoresistance during upwards field sweeps (-10 kOe → 10 kOe) with small angular variations around θ = 90° showing both experimental (a) and simulation data (b). Note that the features are inverted above and below θ = 90°, and that the results are well reproduced by the simulations, as described in the text. Deconstructing the simulation data into contributions from the nanowire legs (green dashed line) and vertices (green dotted line) reveals the critical importance of the vertices to the overall magnetoresistance.

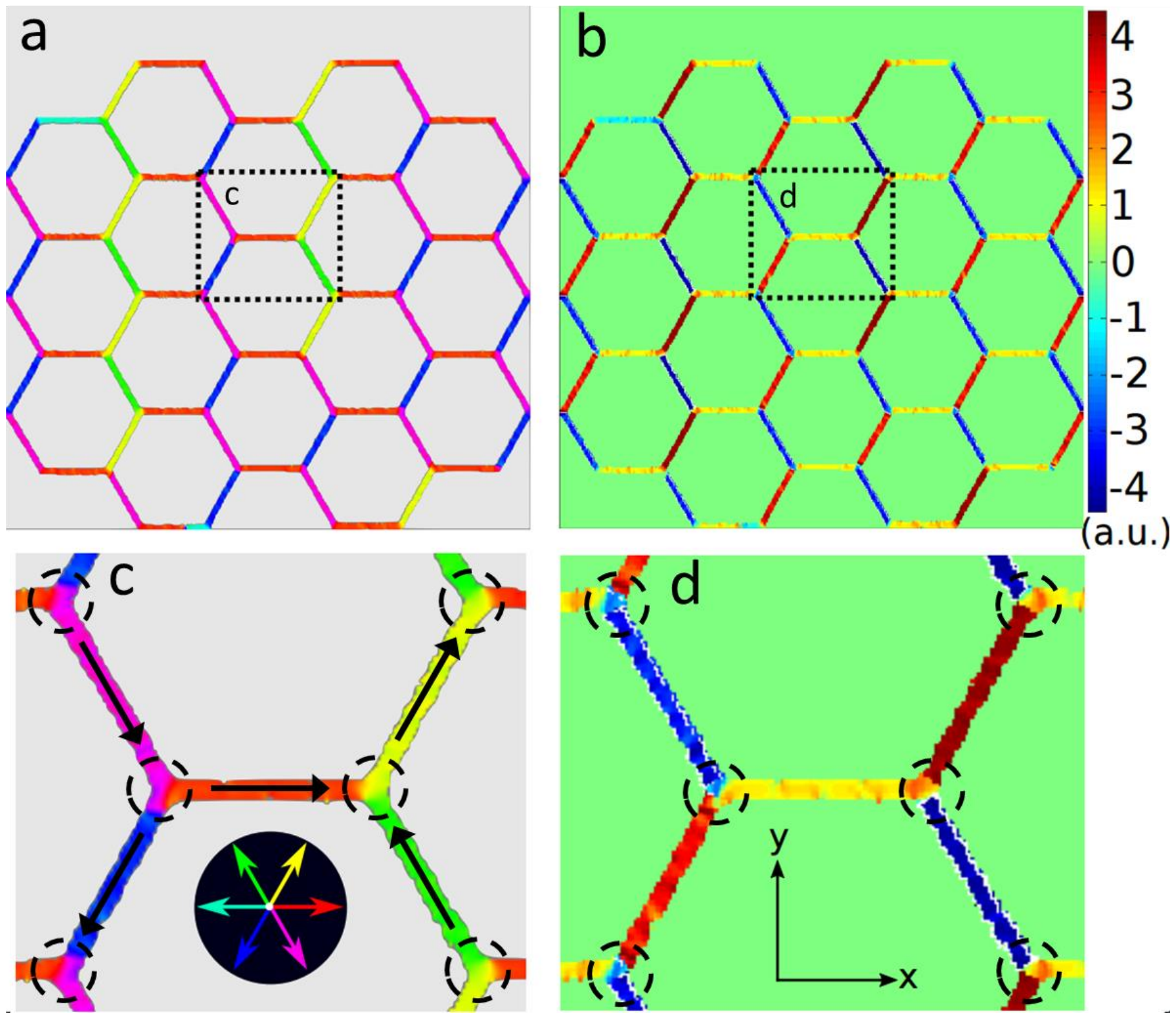

**Figure 5:** Simulated magnetization and y-component electric field maps of a 17-hexagon armchair network. (a) The micromagnetic state at 800 Oe and θ = 90.2° after applying a -4000 Oe saturating field at the same orientation. At this field, only a fraction of the nanowire leg moments have undergone magnetization reversal. (b) Electric field maps of the same state, generated as described in the text. (c) Expanded section of the magnetization map, showing that the vertex regions (circled) have different magnetization profiles that depend on the adjacent nanowire leg moments. (d) Expanded section of the electric field map. Note that the electric field profile of the vertex regions changes with the magnetization profile, while there is no change for the nanowire legs.

**Supplemental Section 1: θ ~ 0° and θ ~ 90° for Armchair and Zigzag Orientations**

In the main text, we presented magnetoresistance (MR) data for armchair-orientation connected kagome artificial spin ice networks. We also focused on the sensitivity of the MR around θ = 90° for armchair-orientation connected kagome artificial spin ice networks. The inversion of the MR features presented in the main text also occurred around θ = 0° for armchair-orientation samples, as well as around θ = 0° and θ = 90° for zigzag-orientation samples. The zigzag design (Fig. S1) is a 90° rotation of the armchair design.

To visualize all the field sweeps simultaneously, we tracked and plotted the transverse magnetoresistance response of a given field strength across every angle. Figure S2 shows this composite curve for various field strengths for both an armchair network and a zigzag network. Notably, the 10 kOe traces behave roughly like $\sin^2$ waveforms, similar to the planar Hall effect response of a thin film. In the armchair device, intermediate field strengths have maxima at -30°/-210° and minima at 30°/-150°, directions which are normal to the 60° and -60° nanowires, respectively. Similarly, the zigzag device displays intermediate field maxima at -240°/-60° and minima at -120°/60°, normal to the 30° and -30° nanowires, respectively. For each device, the zero-field composite curve reveals three distinct values depending on the orientation of the saturating magnetic field, suggesting that the exact micromagnetic configuration has an effect on the zero-field resistance.

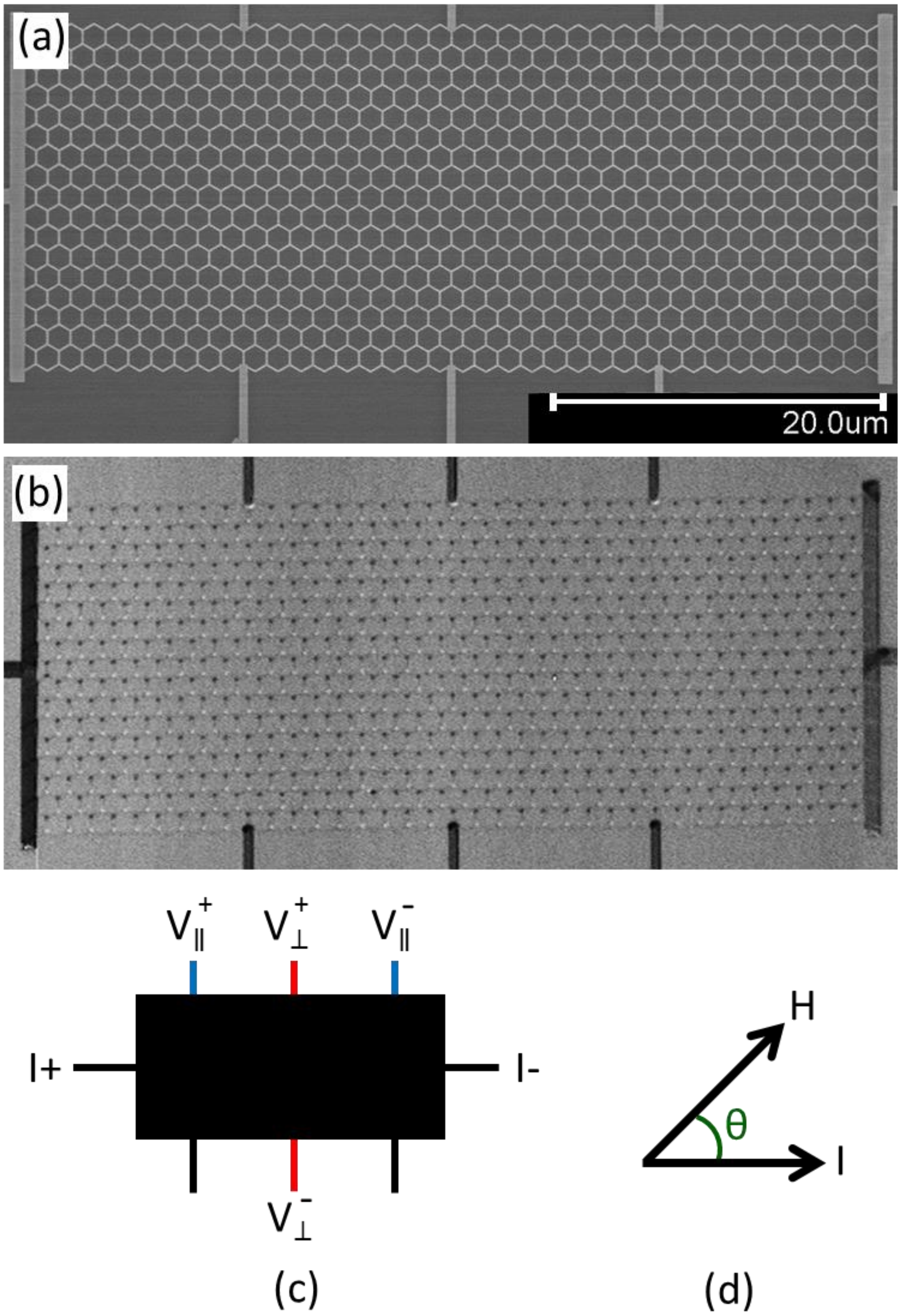

**Figure S1**: (a) SEM image of a zigzag orientation connected kagome artificial spin ice lattice. (b) Corresponding MFM image. The black and white dots at the vertices are indicative of the Ising-like behavior of the individual nanowires. (c) Similar to the armchair orientation networks, large connective pads on each end are used to supply an excitation current while the thin nanowire leads along the long axis are used for voltage measurement. (d) An external magnetic field can be applied along any in-plane direction, denoted by the angle θ between the field direction and the nominal current direction.

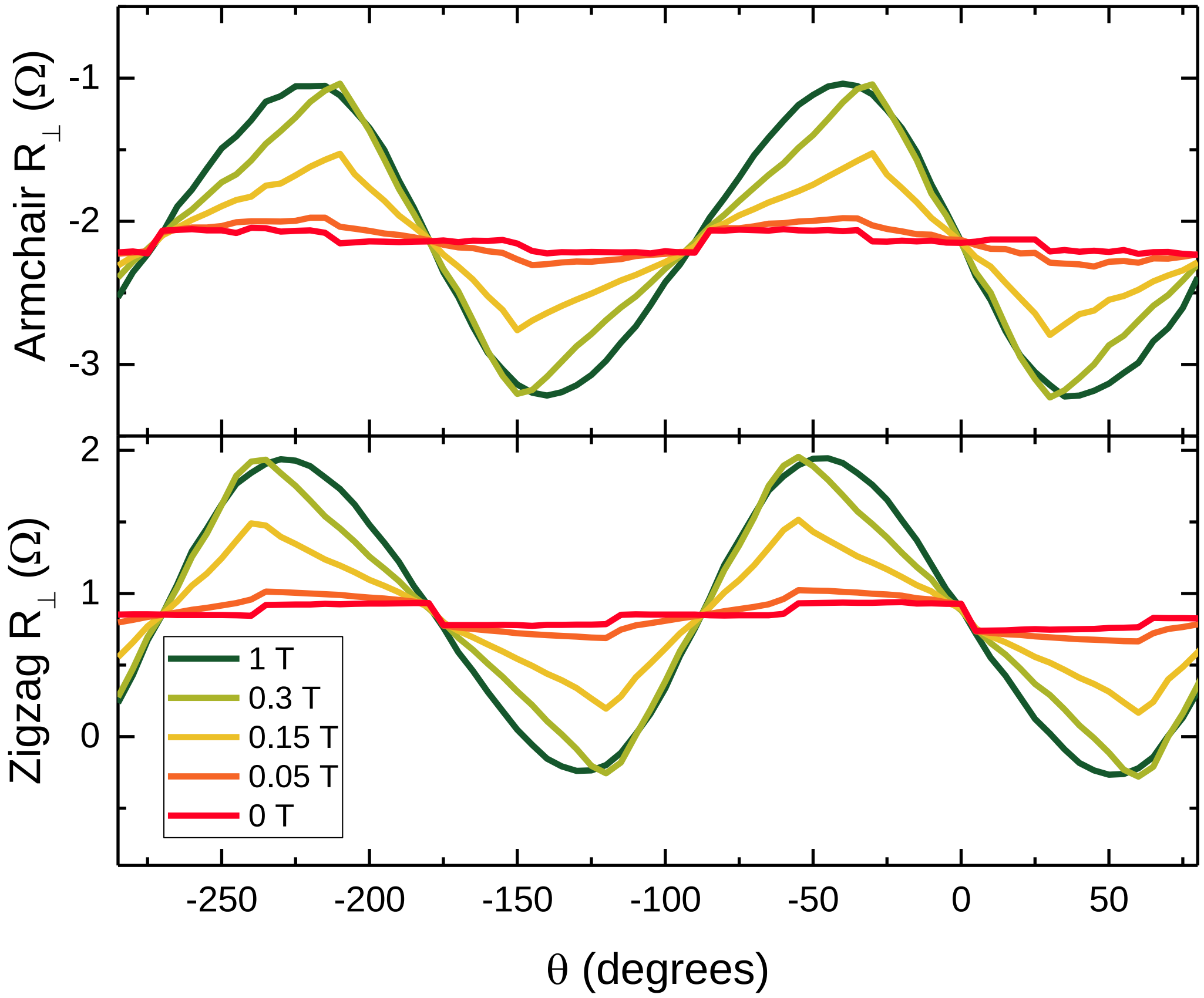

**Figure S2**: For θ from 85° to -290° in 5° increments, the transport data under a hysteretic field sweep were recorded for an armchair device (above) and a zigzag device (below). The above $R_\perp$ vs. θ plots are partial summaries of those data. The points in one of the lines above are not measured successively. For a single curve, the connected data are pieced together from the same field step across every field sweep (downwards sweep). These pseudo-angular plots reveal interesting trends in the data, including varying zero-field resistance. In both devices, the transverse magnetoresistance response is minimized at two sets of pinch points: 0°/180° and ±90°.

The response of the magnetoresistance to a changing external field is minimized at four angles: $\theta \in \{-270°, -180°, -90°, 0°\}$. These 'pinch points' can be easily discerned from the pseudo-angular composite plot and exhibit interesting behavior upon closer inspection. The responses of the armchair and zigzag networks around θ = 90° and θ = 0°, respectively, are qualitatively similar. While the exact details of the transport geometries vary, the orientation of the magnetic field with respect to the lattice is the same in both cases. 90° armchair and 0° zigzag data are presented in Figure S3. In the armchair sample, the magnetoresistance responses mirror about the x-axis as the angle crosses 90°. This change in slope is most drastic at approximately 400 Oe and 1700 Oe. In the zigzag sample, the change in slope occurs as the angle crosses 0° and is most noticeable at approximately 450 Oe and 1900 Oe.

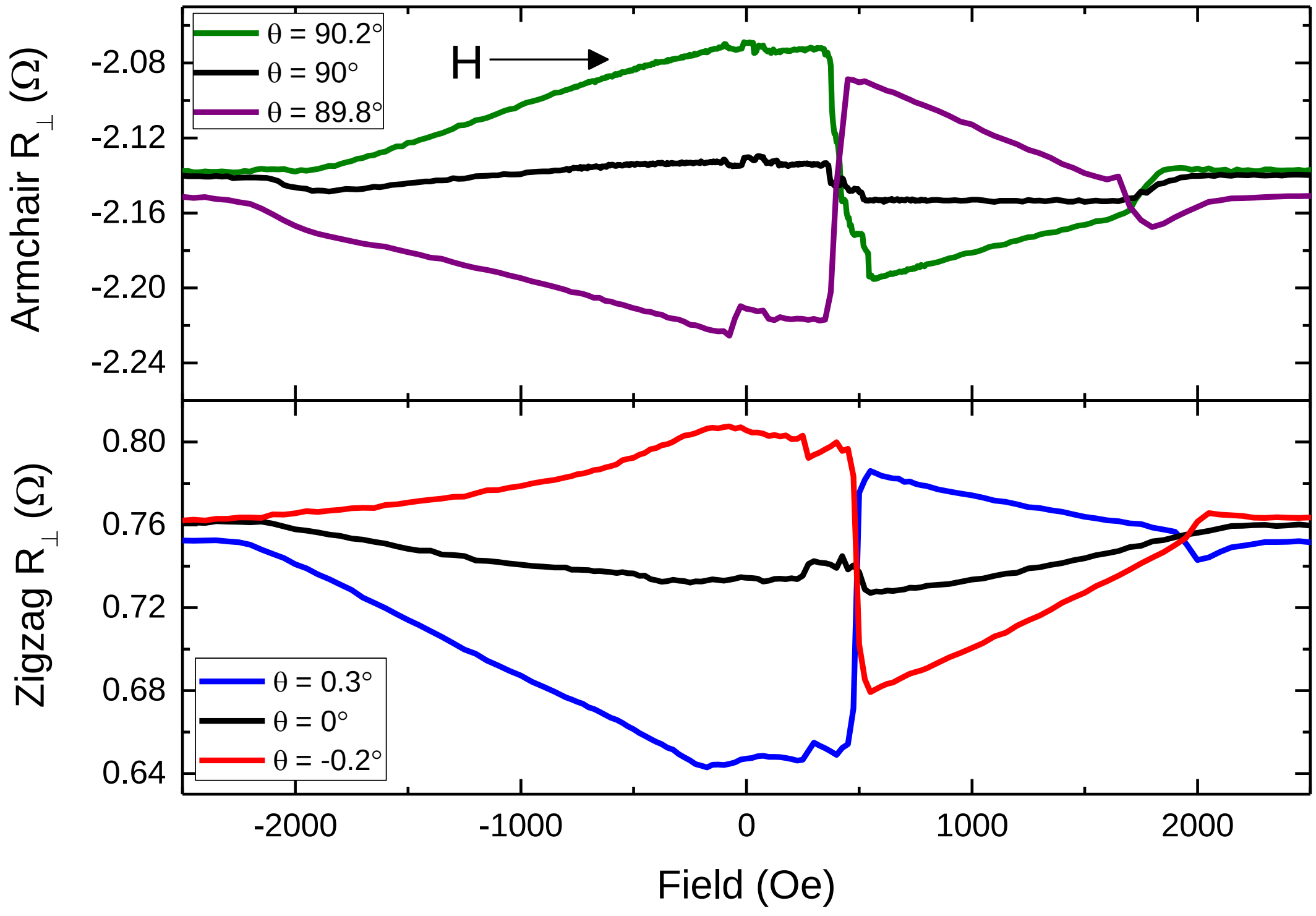

**Figure S3**: Field sweep measurements with small angular variations around (above) θ = 90° for an armchair device and (below) θ = 0° for a zigzag device. As the applied field angle crossed 0° or 90°, the plot features flipped about the x-axis.

In Figure S4, we also grouped the 0° armchair and 90° zigzag data due to their similarities. In the armchair orientation, for field sweeps with θ near 0°, we again observed that small angular changes to the applied magnetic field resulted in large effects on the magnetoresistance response, with the response inverting as the angle crossed 0° (Fig. 4b). This effect was most noticeable in a sharp resistance peak (θ < 0°) or valley (θ > 0°) at approximately 550 Oe. The resistance also showed a mirroring feature near -300 Oe, although it should be noted that no magnetization reversal was expected as the field approaches zero from saturation at -10 kOe. In the zigzag orientation, the peak/valley flip occurred at approximately the same field value (~600 Oe), but no bump near -300 Oe was observed.

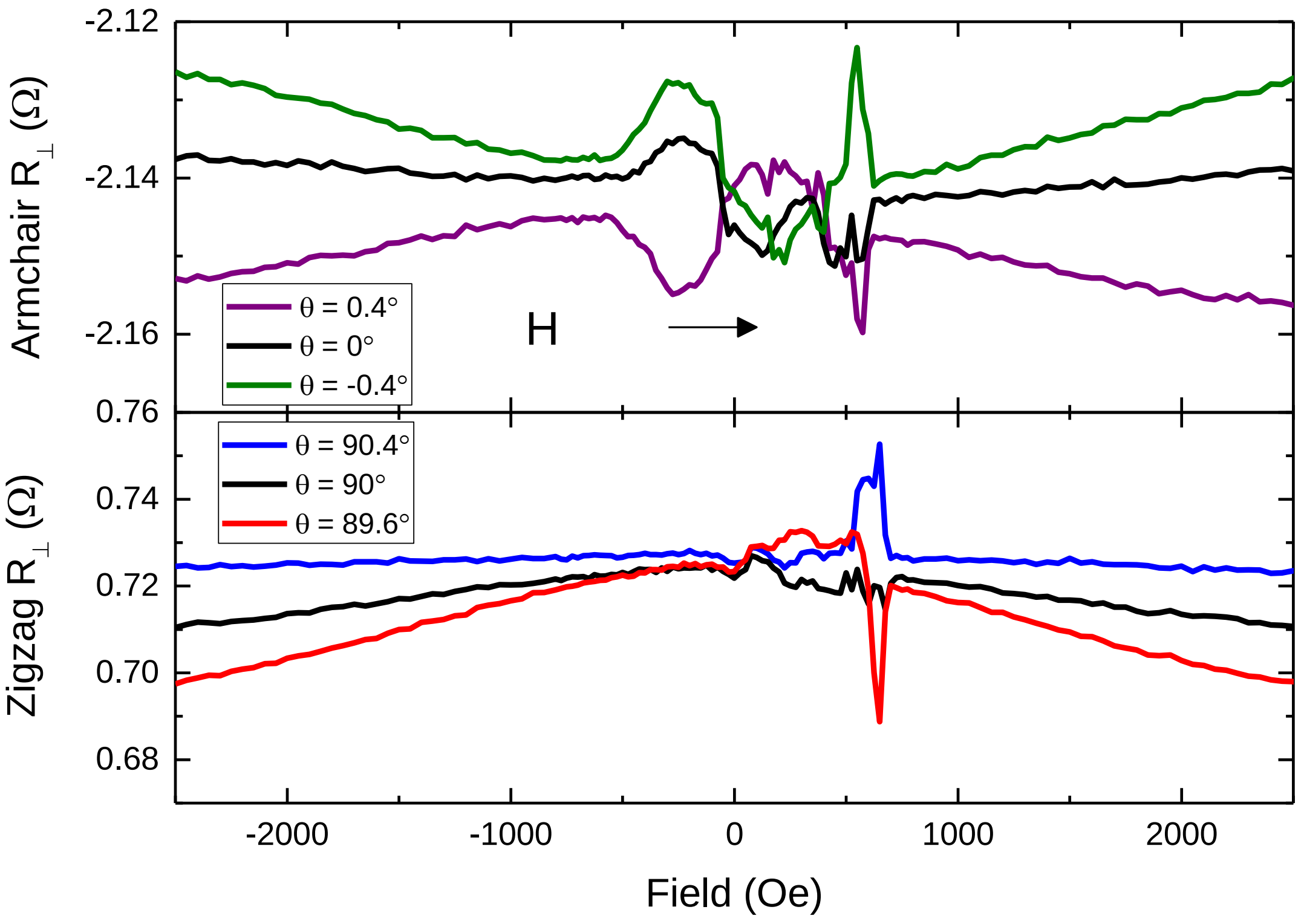

**Figure S4**: Field sweeps with small angular variations around (above) θ = 0° for an armchair device and (below) θ = 90° for a zigzag device. As the applied field angle crossed 0° or 90°, the plot features flipped about the x-axis.

## Supplemental Section 2: Further Comparisons of Experimental and Simulated MR

Here we present experimental magnetoresistance (MR) measurements for a second device in the armchair configuration. Figure S5 compares them to the same simulation data as in Figure 2 at the same applied field angles (0°, 30°, 60°, and 90°). Figure S6 compares experimental and simulated MR at three other angles (15°, 45°, 75°).

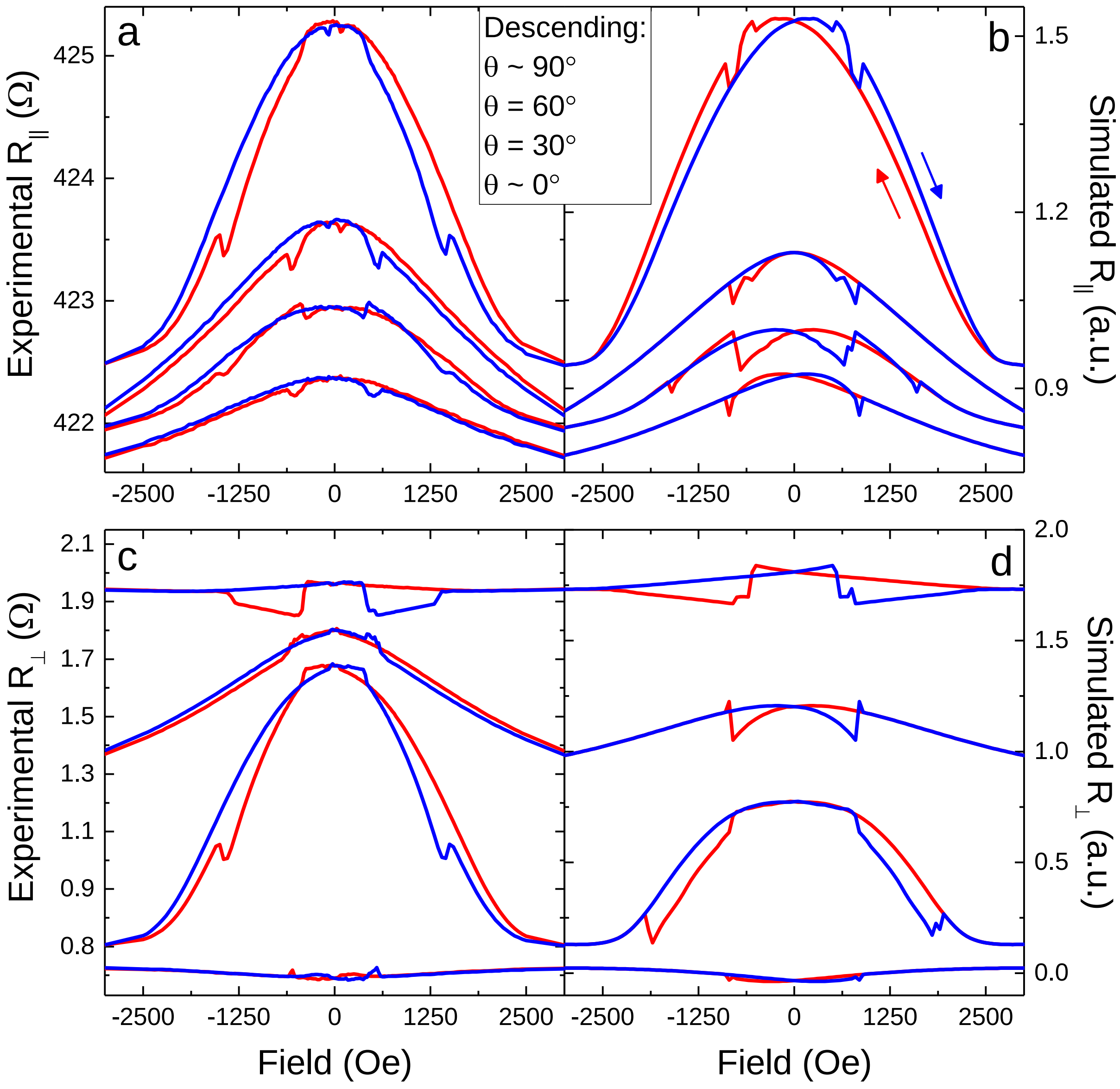

**Figure S5**: Comparison of experimental MR (left) and simulated MR (right) for the second armchair-orientation device. The experimental results are consistent with those of the first device.

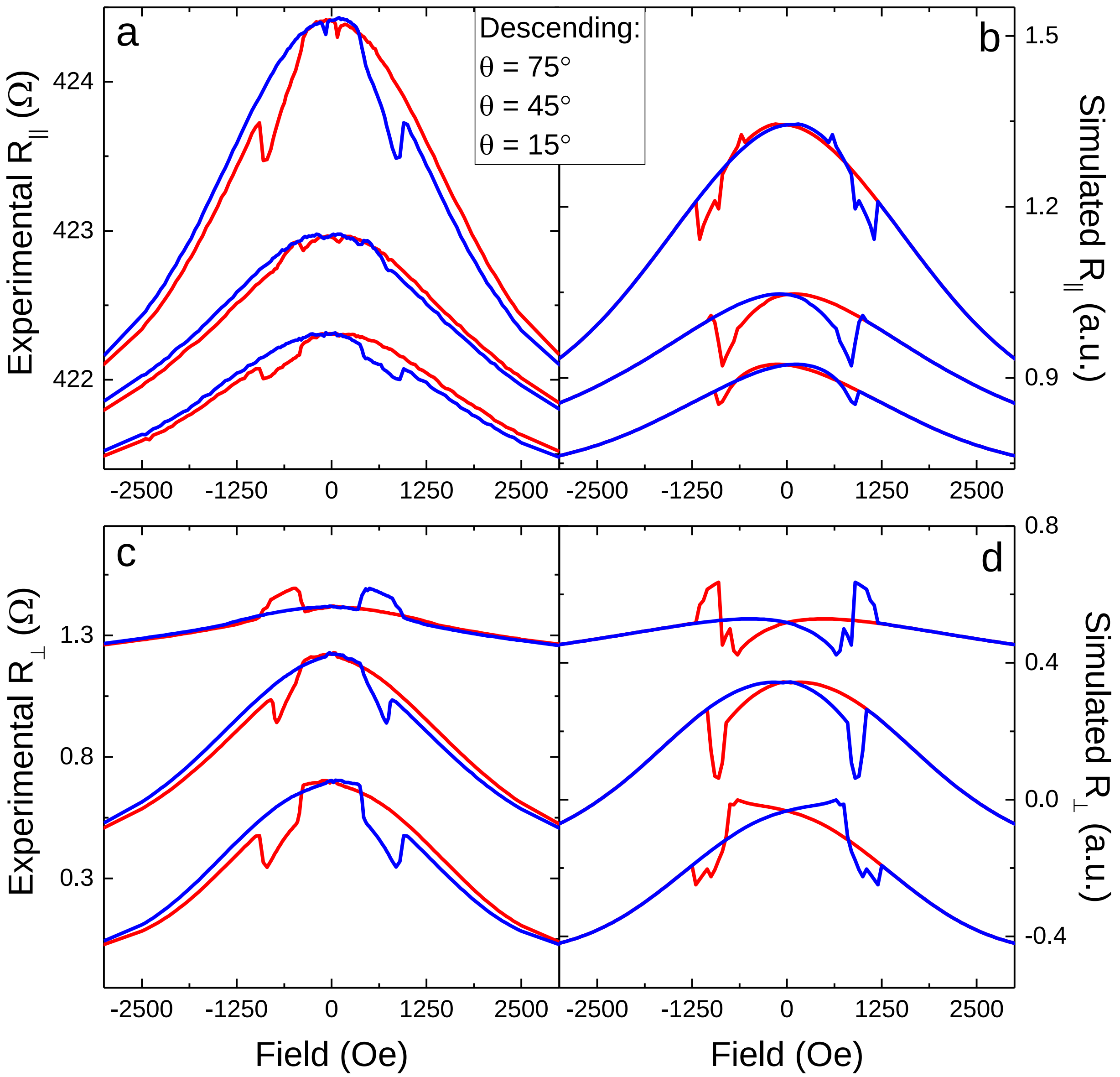

**Figure S6**: Comparison of experimental MR and simulated MR for the second armchair device for field angles of 15°, 45°, and 75°. The overall features of the experimental data are captured fairly well, with the exception of the 45° longitudinal data.

**Supplemental Section 3: Mumax3 Micromagnetic Modeling and MR Extraction**

A 512 x 512 pixel bitmap image of the armchair lattice use in the Mumax3 simulations is shown below in Fig. S7. The image was generated from a scanning electron micrograph of an experimental device. Each pixel corresponds to a 12.5 x 12.5 nm$^2$ micromagnetic cell. The volume of each micromagnetic cell was 12.5 x 12.5 x 6.25 nm$^3$. The electric field in each cell is calculated using the expression:

$$\mathbf{E} = \rho_0 \mathbf{J} + \Delta\rho(\hat{\mathbf{m}} \cdot \mathbf{J})\hat{\mathbf{m}} \tag{S1}$$

Here the quantity $\rho_0$ is the bulk resistivity and $\Delta\rho$ is the contribution from anisotropic magnetoresistance. Throughout the simulation, we set $\rho_0 = 1$ and $\Delta\rho = 0.05$. Once the electric field values are obtained we can model the experimental results by integrating the electric field along paths that we expect to mimic the longitudinal and transverse signals. Figure S7 also has four points marked on it (1-4). We obtained voltages from the simulations by integrating the electric field along representative paths mimicking lead placements in actual devices:

$$V_{long} = -\int_{1}^{2} \mathbf{E} \cdot \mathrm{d}\mathbf{s} \tag{S2}$$

$$V_{trans} = -\int_{3}^{4} \mathbf{E} \cdot \mathrm{d}\mathbf{s} \tag{S3}$$

Points 1 and 2 are the respective start and stop points of the longitudinal path over which the electric field was integrated. Similarly, points 3 and 4 are the respective start and stop points of the transverse path. Integrating over different paths (but with the same start and stop points) did not significant change the results.

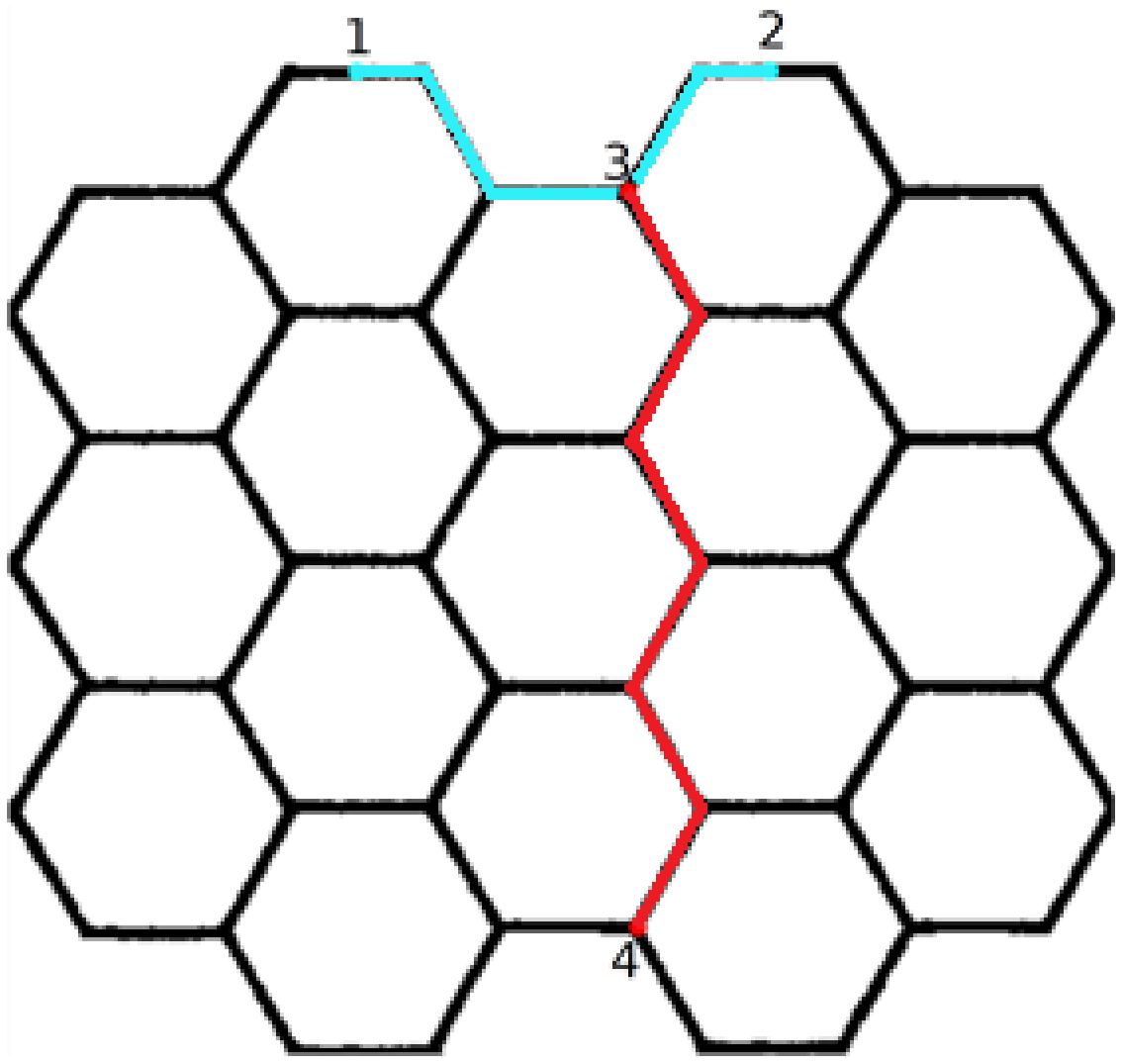

**Figure S7**: The bitmap image used for micromagnetic simulations is shown above. The teal path between points 1-2 is the longitudinal path over which the electric field was integrated, while the red path between 3-4 is the transverse path.

When calculating the electric field for each micromagnetic map we use a 0$^{th}$ order map of the current density as described in the main body of the text. For convenience we illustrate a unit of the 0$^{th}$ order map in Fig. S8c. In principle, higher order corrections in both $\mathbf{E}$ and $\mathbf{J}$ can be obtained by iteratively using Eq. (1). For example, the first-order correction to the current is simply $\mathbf{J}^{(1)} = \mathbf{E}^{(1)}/\rho_0$. Here in Fig. S8a and Fig. S8b we show the x- and y-component of the current density that we used for calculation of the electric field.

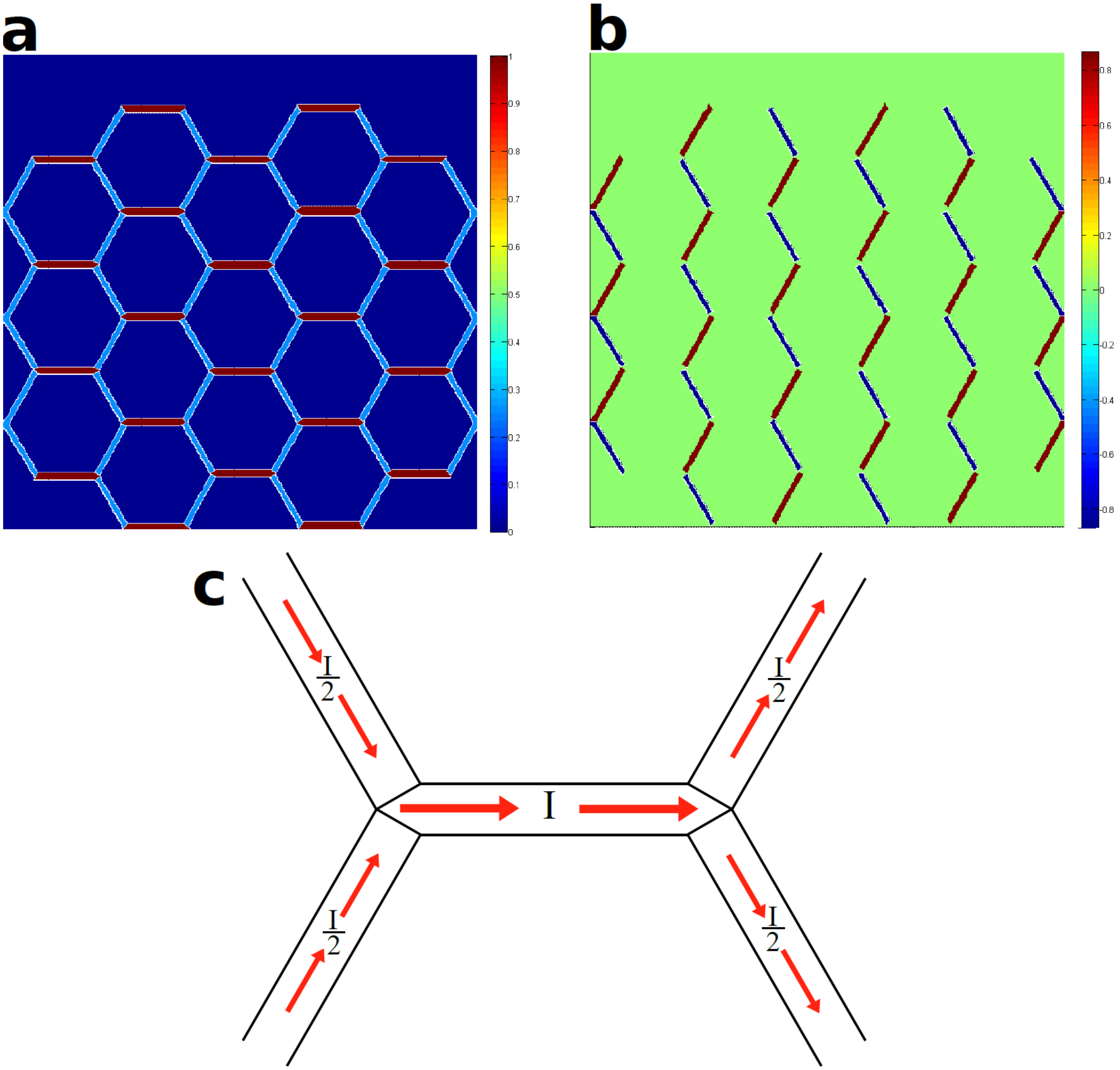

**Figure S8**: The x- and y-components of the current density through the network are shown in (a) and (b) respectively.

In Fig. S9 we present additional micromagnetic and x- and y-component electric field maps of a simulation where the system began magnetized with a field of magnitude -4000 Oe at θ = 90.2°. The field was then reversed in steps of 50 Oe to a configuration of 4000 Oe at θ = 90.2°. This configuration corresponds to a state where the transverse signal arises from vertex effects alone as seen in MR sweeps around this angle (Figures 4 and 6 in main text).

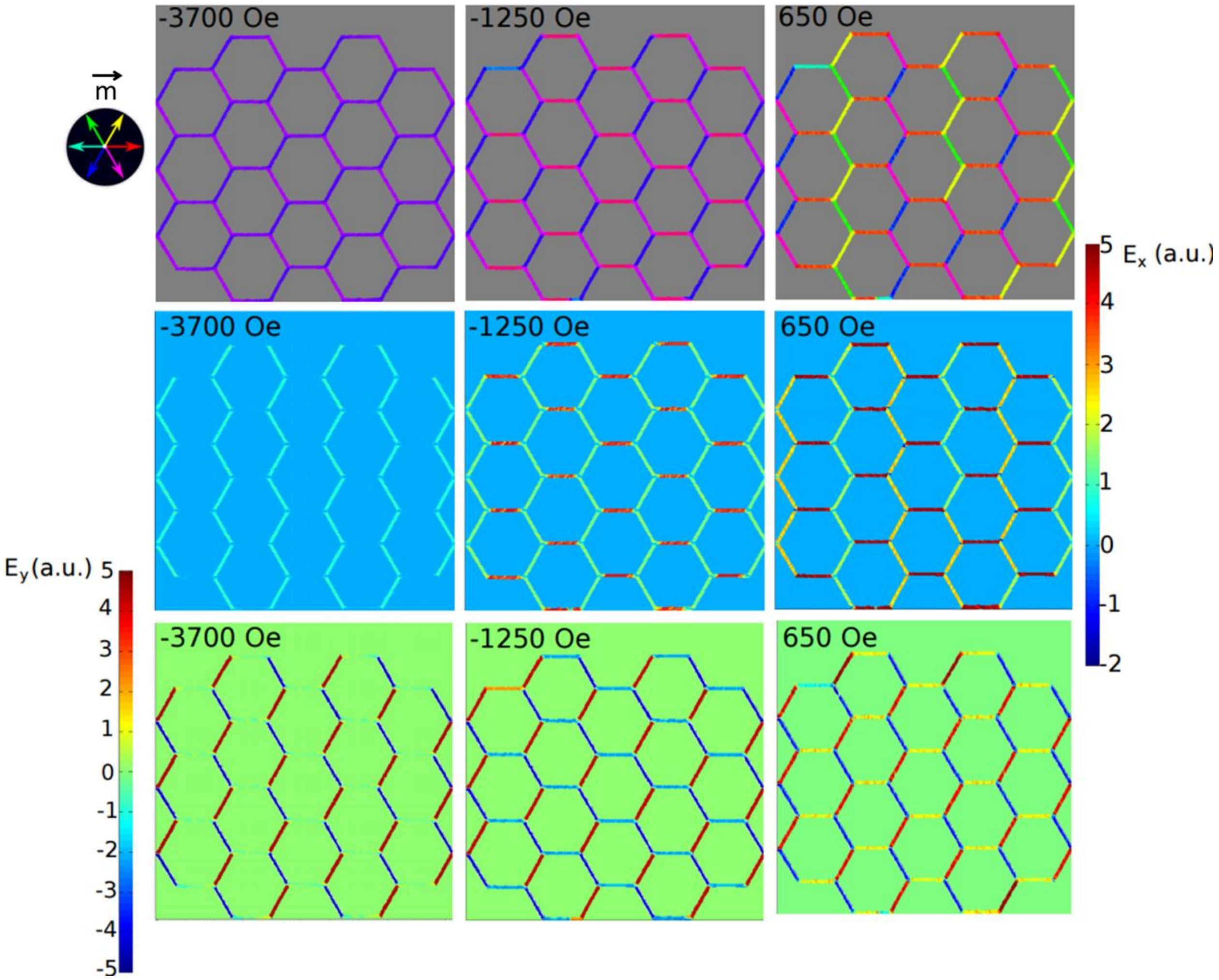

**Figure S9**: Additional micromagnetic maps (top row), x-component of electric field maps (middle row), and y-component of electric field maps (bottom row) at θ = 90.2°

During an experimental field sweep the transverse voltage measured is not centered around zero, instead it is centered about a small negative voltage. Most likely this small offset is due to a longitudinal signature being picked up by slight misalignment of the leads. When performing simulations, in principle, we should have no offset problem and the transverse signal should be centered around zero. However, an offset can artificially occur if the finite elements used in the numeric integration are not sampled symmetrically.  Specifically, if more elements are integrated in region 1 compared to region 4 in Fig. S10 there will be an artificial offset (the same statement can be made with regards to region 2 and 3).  Because our micromagnetic map was drawn from an SEM image and further coarse-grained by MuMax3 the number of finite elements in a given region does not exactly match that region's counterpart.  The number of finite elements between regions 1 and 4 and 2 and 3 is within 5% of each other; but because we work in arbitrary units this leads to a misleading offset in our transverse magnetoresistance curves.  To account for this numeric issue, when calculating the line integral we keep track of the four regions separately and normalize as follows:

$$V_{Trans} = \left[\frac{S_1}{N_1} + \frac{S_4}{N_4}\right] \times [N_1 + N_4] + \left[\frac{S_2}{N_2} + \frac{S_3}{N_3}\right] \times [N_2 + N_3]$$

For example, here $S_1$ is the contribution to the line integral from the 1st region and $N_1$ is the number of finite elements in the 1st region.

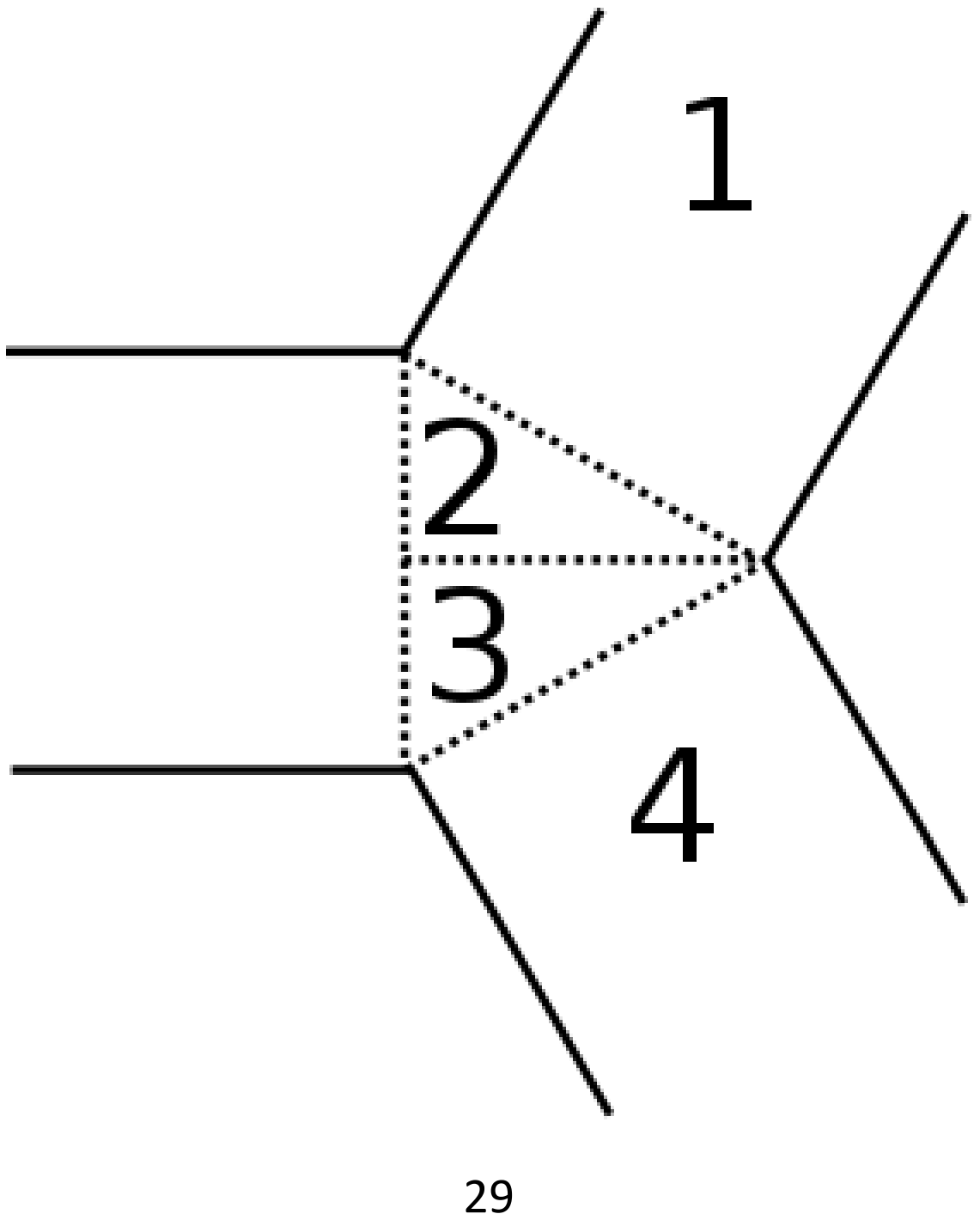

**Figure S10**: When calculating the transverse line integral we keep track of the contribution to the line integral from the four regions illustrated above separately. By doing this we apply a normalization procedure (outlined above) that allows us to avoid unphysical offsets in the transverse voltage.

It is instructive to also look at the contributions to the line integral from region 1 and 4 in comparison to region 2 and 3. In the main text we call this separation the leg region and the vertex region respectively. As shown in Figure 4 within the main text there are certain conditions where the vertex region dominates the transverse MR relative to the leg region. In Figure S11 we show additional angles where the simulated MR is broken into these two parts. For intermediate angles such as θ = 60° and θ = 30° the leg region is seen to contribute more to the MR over the entire field range.

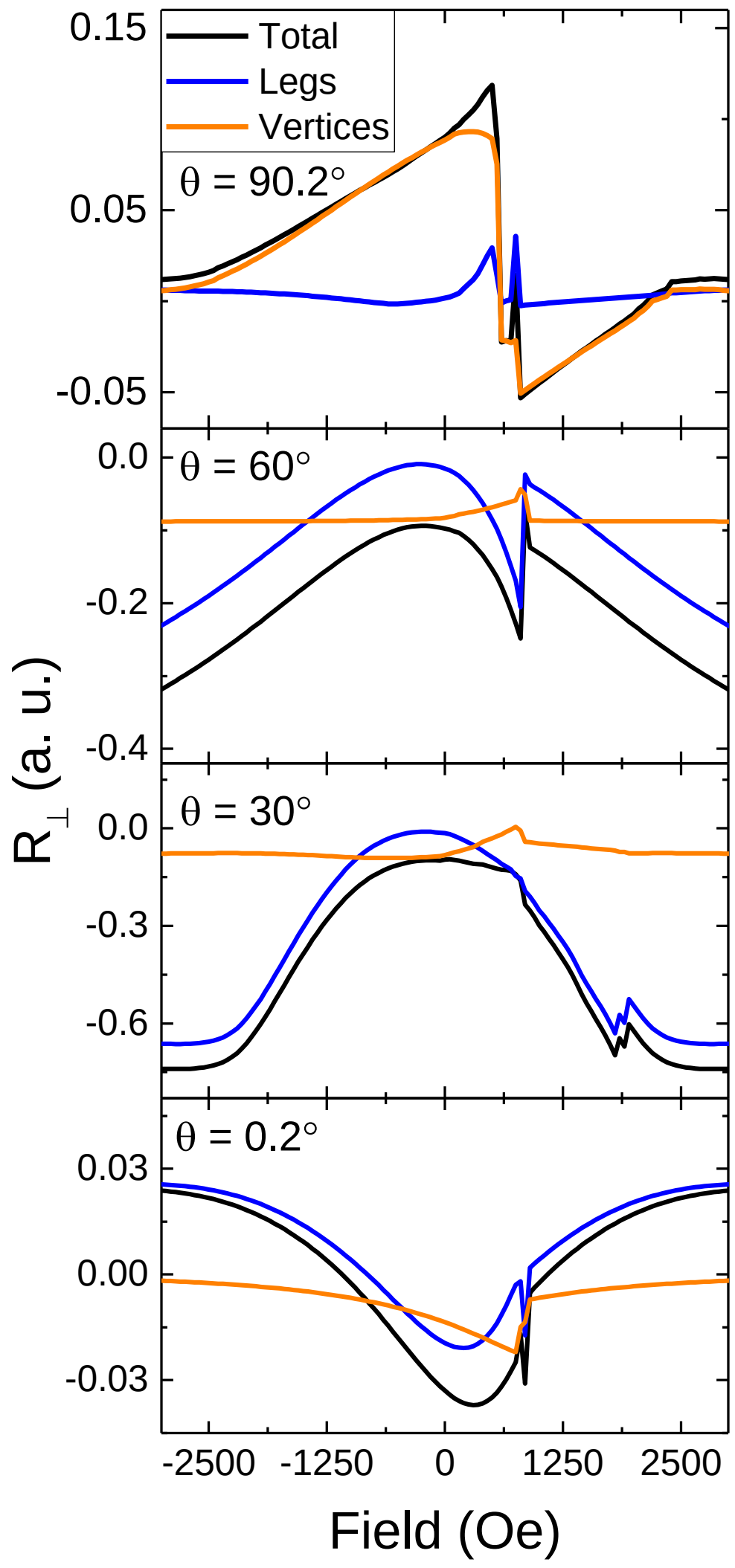

**Figure S11**: Simulated transverse magnetoresistance signals (black) and separated contributions from the nanowire legs (blue) and the vertices (orange), taken by simply separating different parts of the line integrals. The magnetic field was swept from -4000 Oe to +4000 Oe. The significant contributions of the vertex regions are clearly evident near 90 degrees and zero degrees. We note that, for all angles, near 0 Oe the contribution from the legs is approximately zero as is expected in the Ising limit. The leg contribution does not completely vanish because our definition of the vertex region is somewhat smaller than the region in which the vertex impacts the magnetization profile.

**Supplemental Section 4: Sample Fabrication and Measurement Details**

Samples were patterned *via* electron-beam lithography on a $Si_3N_4$ coated (~200 nm) silicon substrate. Permalloy ($Ni_{81}Fe_{19}$) was deposited *via* ultra-high vacuum electron-beam evaporation from an alloy target at 0.5 Å/s in a system with a base pressure in the $10^{-10}$ Torr range. Roughly 41 hexagons span the long axis with 15 hexagons across the short axis (636 total hexagons, including 32 partial hexagons on the ends). Individual legs of the permalloy network were approximately 800 nm by 75 nm in-plane and 25 nm thick. Samples with permalloy thicknesses of 12 nm and 40 nm were also fabricated and resulted in qualitatively similar measurement results compared to the 25 nm thick permalloy samples, strongly suggesting that the type of domain wall in the system (transverse vs. vortex, see reference [29] in the main text) does not significantly affect the results.

Connected artificial spin ice samples were mounted on a rotating platform inside a cryostat with a fixed-direction ±14 T magnet. The sample was rotated such that the magnetic field orientation was always parallel to the sample surface (in-plane), with θ varying from -280° to 100°. Magnetotransport measurements were taken at room temperature (constant 300 K regulated in the cryostat) using lock-in amplifiers with a 17 Hz ac excitation ($I$ = 66.7 μA) applied across the long axis of the sample. No qualitative differences were seen when varying the excitation frequency (17 Hz – 1700 Hz) or current amplitude (7.42 μA – 66.7 μA). The magnetoresistance responses of four separate networks (two armchair geometry, two zigzag geometry) were taken after initial MFM characterization, and the results were qualitatively consistent among the different samples.